\documentclass[useAMS,usenatbib]{mn2e}
\usepackage{graphics,rotate}
\def\gtsim{\mathrel{\hbox{\rlap{\hbox{\lower4pt\hbox{$\sim$}}}\hbox{$>$}}}}
\def\lesssim{\mathrel{\hbox{\rlap{\hbox{\lower4pt\hbox{$\sim$}}}\hbox{$<$}}}}

\def\etal{{\it et al.\ }}
\def\eg{{\it e.g.\ }}

\def\spose#1{\hbox to 0pt{#1\hss}}
\def\approxlt{\mathrel{\spose{\lower 3pt\hbox{$\sim$}}
        \raise 2.0pt\hbox{$<$}}}
\def\approxgt{\mathrel{\spose{\lower 3pt\hbox{$\sim$}}
        \raise 2.0pt\hbox{$>$}}}
\def\approxpropto{\mathrel{\spose{\lower 3pt\hbox{$\sim$}}
        \raise 2.0pt\hbox{$\propto$}}}
\mathchardef\twiddle="2218

\def\multleft#1{\hbox to size{\vbox {\halign {\lft{##}\cr #1}}\hfill}\par}
\def\multright#1{\hbox to size{\vbox {\halign {\rt{##}\cr #1}}\hfill}\par}

\def\today{\ifcase\month\or January\or February\or March\or April\or May\or
      June\or July\or August\or September\or October\or November\or December\fi
      \space\number\day, \number\year}
\def\<{\thinspace}

\def\km{{\rm\thinspace km}}

\def\Mpc{{\rm\thinspace Mpc}}
\def\Msun{\hbox{$\rm\thinspace M_{\odot}$}}

\def\s{{\rm\thinspace s}}

\def\kmps{\hbox{$\km\s^{-1}\,$}}

\def\kmpspMpc{\hbox{$\kmps\Mpc^{-1}$}}

\title[The dark matter halos of relaxed galaxy clusters]
{The dark matter halos of massive, relaxed galaxy clusters observed 
with Chandra}

\author[Schmidt \& Allen]
        {R. W. Schmidt$^1$\thanks{E-mail: rschmidt@ari.uni-heidelberg.de}
        and S. W. Allen$^2$\\
        1. Astronomisches Rechen-Institut, Zentrum f{\"u}r
        Astronomie der Universit{\"a}t Heidelberg, M{\"o}nchhofstrasse
        12-14, 69120
	Heidelberg, Germany\\
        2. Kavli Institute for Particle Astrophysics and Cosmology,
	Stanford University, 382 Via Pueblo Mall, Stanford, CA
	94305-4060, USA
}
\date{Draft \today 
}

\pagerange{\pageref{firstpage}--\pageref{lastpage}}
\pubyear{2006}

\begin{document}

\maketitle

\label{firstpage}

\begin{abstract}
We use the Chandra X-ray Observatory to study the dark matter halos of
34 massive, dynamically relaxed galaxy clusters, spanning the redshift
range $0.06<z<0.7$. The observed dark matter and total mass
(dark-plus-luminous matter) profiles can be approximated by the
Navarro Frenk \& White (hereafter NFW) model for cold dark matter
(CDM) halos; for $\sim 80$ per cent of the clusters, the NFW model
provides a statistically acceptable fit. In contrast, the singular
isothermal sphere model can, in almost every case, be completely ruled
out. We observe a well-defined mass-concentration relation for the
clusters with an
intrinsic scatter in good agreement
with the predictions from simulations. The slope of the
mass-concentration relation, $c\propto M_{\rm vir}^a/(1+z)^b$ with
$a=-0.45\pm0.12$ at 95 per cent confidence, is steeper than the value
$a\sim -0.1$ predicted by CDM simulations for lower mass halos. With
the slope $a$ included as a free fit parameter, the redshift evolution
of the concentration parameter, $b=0.71\pm0.52$ at 95 per cent
confidence, is consistent with the same simulations ($b\sim1$).
Fixing $a\sim -0.1$ leads to an apparent evolution that is
significantly slower, $b=0.30\pm0.49$, although the goodness of fit in
this case is significantly worse. Using a generalized NFW model, we
find the inner dark matter density slope, $\alpha$, to be consistent
with unity at 95 per cent confidence for the majority of
clusters. Combining the results for all clusters for which the
generalized NFW model provides a good description of the data, we
measure $\alpha=0.88\pm0.29$ at 95 per cent confidence, in agreement
with CDM model predictions.
\end{abstract}


\section{Introduction}
\label{section:intro}

One of the most remarkable results of cold dark matter (CDM)
simulations of structure formation is that the density profiles of
dark matter halos on all resolvable mass scales, from small satellites
to the most massive galaxy clusters, can be approximated by a universal
profile, the so-called Navarro-Frenk-White (NFW) profile \citep*{Navarro95,Navarro97}

\begin{equation}
\rho(r)=\frac{\rho_0}{(\frac{r}{r_{\rm s}})
(1+\frac{r}{r_{\rm s}})^{2}}.
\label{nfwprofile}
\end{equation}

\noindent Here $r_{\rm s}$ is the characteristic scale radius of the halo and 
$\rho_0$ is the central density. The model has a central, negative 
logarithmic density slope, referred to hereafter as the ``inner
slope''

\begin{equation}
\alpha=\left.-\frac{{\rm d\,log}\rho}{{\rm d\,log}r}\right|_{r\rightarrow 0}=1.
\end{equation}
NFW also defined the concentration parameter, $c$, as the ratio of 
$r_{200}$, the radius within which
the matter density is 200 times the critical density, and the scale
radius:

\begin{equation}
c=\frac{r_{\rm 200}}{r_{\rm s}}.
\end{equation}
They showed that smaller mass halos are more concentrated than the
higher mass halos and interpreted this as reflecting the higher
formation redshift of the lower mass systems.

Copious numerical work has been devoted to testing these fundamental
findings. To explain the mass-concentration relation and its redshift
evolution, \citet{Bullock01} and \citet{Eke01} introduced simple, but highly
successful models for the formation of dark matter halos.  In 
the \citet{Bullock01} model, for example,
only two parameters, $K$, which determines the initial
concentration parameter of collapsing halos and $F$, the ratio of the
initial collapse mass to the final virial mass of the halo at redshift
zero, are required to approximately match the simulation
predictions. More recently simple two-parameter power-laws have been
used to characterize the mass-concentration relation
\citep{Dolag04,Shaw05}.

When considering the inner slopes of dark matter halos, a useful
generalization of the NFW formula is

\begin{equation}
\rho(r)=\frac{\rho_0}{(\frac{r}{r_{\rm s}})^{\alpha}
(1+\frac{r}{r_{\rm s}})^{3-\alpha}}
\label{gnfw}
\end{equation}
\citep[e.g., ][]{Hernquist90,Zhao96,Jing00}, where the inner slope
$\alpha$ is a free parameter. From their initial simulations,
\citet{Moore99} found a steeper central slope than NFW, with
$\alpha=1.5$. For some years, the question of the precise value for
the inner slope remained the topic of much debate. However, a
consistent view has now emerged in which real dispersion is expected
between the inner slopes for individual halos
\citep{Klypin01,Tasitsiomi04} and where typical values for the inner
slopes of clusters lie in the range $\alpha\sim1.1\pm0.4$
\citep{Navarro04,Diemand04,Diemand05}.
Recent analytical and numerical work to solve the Jeans equation
suggests a value for the inner slope of $\alpha\approx 0.8$
\citep{Austin05,Dehnen05,Hansen06}.

In summary, both a tight mass-concentration relation (for recent
results see e.g., \citealt{Shaw05}) and an inner density slope for
dark matter halos in the range $0.7<\alpha<1.5$ are central
predictions of the CDM paradigm for structure formation.

In this paper we use Chandra X-ray Observatory data to study the mass
profiles for 34 of the most massive, dynamically relaxed galaxy
clusters known. Such clusters are among the most promising targets
with which to check the central CDM predictions, being both dominated
by dark matter \citep[e.g., ][and references therein]{Allen04} and having
a size that allows us to resolve well within the scale radius, even at
high redshifts.  We employ an analysis method which minimizes the need
for priors associated with the use of parameterized models for
the X-ray gas density and/or temperature profiles. The restriction to
relaxed clusters also minimizes systematic effects associated with
\eg geometry and possible non-thermal pressure support.

Previous X-ray studies have suggested broad agreement with the
theoretically predicted mass-concentration relation
\citep{Pointecouteau05,Zhang06,Vikhlinin06,Voigt06}, at least at low
redshifts. Here, we increase the detail of this comparison and, for
the first time, measure both the slope of the mass-concentration
relation and its evolution. We show that tension may exist between the
Chandra data and some simple models based on CDM simulations for lower
mass halos.  With regard to the inner density slope, Chandra and
XMM-Newton results to date have, in general, suggested good agreement
with the CDM predictions
\citep{Allen01a,Schmidt01,Allen02,Lewis02,Lewis03,Buote04,Arabadjis02,Arabadjis04,Andersson04}. After
some initial controversy, there is also an emerging consensus that
strong gravitational lensing data support inner dark matter density
slopes of about unity in several clusters
\citep{Smith01,Gavazzi03,Sand02,Sand04,Bartelmann04,Dalal03,Meneghetti05,Sand05}. We
extend this work and show that the NFW model provides a good
description of the total mass and dark matter profiles for most
relaxed clusters, rejecting completely the simple singular isothermal
model ($\rho \propto 1/r^2$). We also obtain a robust result on the
inner slope for the ensemble of clusters.

A flat $\Lambda$CDM reference cosmology is assumed with Hubble
constant $H_0=70$\kmpspMpc and matter density $\Omega_{\rm m}
=0.3$. In a few cases cluster masses are quoted with a Hubble constant
scale $h=H_0/100$\kmpspMpc.

\section{Method}

\begin{table*}
\begin{center}
\caption{Summary of the Chandra observations. Columns list the 
target name, redshift, observation date, detector used, observation mode, 
net exposure after all cleaning and screening processes, and 
coordinates from the X-ray centres used in the analysis. 
Where multiple observations of a single target have been used, 
these are listed separately.  Redshifts for the MACS clusters are
from Ebeling \etal 2006, in preparation, and will appear in full
in the published article.}\label{table:obs}
\vskip 0 truein
\begin{tabular}{ l c c c c c c c c c }
&&&&&&&&&  \\
                     & ~ & z      &~~ &    Date      & Detector & Mode & Exposure (ks) & R.A. (J2000.)  &    DEC. (J2000.)     \\
\hline                                
                                      
Abell 1795(1)        & ~ & 0.063  & &    2002 Jun 10  & ACIS-S & VFAINT & 13.2   &   13 48 52.4  & 26 35 38  \\
Abell 1795(2)        & ~ &  ``    & &    2004 Jan 14  & ACIS-S & VFAINT & 14.3   &      ``       &    ``     \\
Abell 1795(3)        & ~ &  ``    & &    2004 Jan 18  & ACIS-I & VFAINT & 9.6    &      ``       &    ``     \\ 
Abell 2029(1)        & ~ & 0.078  & &    2000 Apr 12  & ACIS-S & FAINT  & 19.2   &   15 10 56.2  & 05 44 41  \\
Abell 2029(2)        & ~ &  ``    & &    2004 Jan 08  & ACIS-S & FAINT  & 74.8   &      ``       &    ``     \\
Abell 2029(3)        & ~ &  ``    & &    2004 Dec 17  & ACIS-I & VFAINT & 9.4    &      ``       &    ``     \\
Abell 478(1)         & ~ & 0.088  & &    2001 Jan 27  & ACIS-S & FAINT  & 39.9   &   04 13 25.2  & 10 27 55  \\
Abell 478(2)         & ~ &  ``    & &    2004 Sep 13  & ACIS-I & VFAINT & 7.4    &      ``       &    ``     \\
PKS0745-191(1)       & ~ & 0.103  & &    2001 Jun 16  & ACIS-S & VFAINT & 17.4   &   07 47 31.7  &-19 17 45  \\
PKS0745-191(2)       & ~ &  ``    & &    2004 Sep 24  & ACIS-I & VFAINT & 9.2    &      ``       &    ``     \\
Abell 1413           & ~ & 0.143  & &    2001 May 16  & ACIS-I & VFAINT & 64.5   &   11 55 18.1  & 23 24 17  \\
Abell 2204(1)        & ~ & 0.152  & &    2000 Jul 29  & ACIS-S &  FAINT & 10.1   &   16 32 47.2  & 05 34 32  \\
Abell 2204(2)        & ~ &  ``    & &    2004 Sep 20  & ACIS-I & VFAINT & 8.5    &      ``       &    ``     \\
Abell 383(1)         & ~ & 0.188  & &    2000 Nov 16  & ACIS-S &  FAINT & 18.0   &   02 48 03.5  &-03 31 45  \\
Abell 383(2)         & ~ &  ``    & &    2000 Nov 16  & ACIS-I & VFAINT & 17.2   &      ``       &    ``     \\
Abell 963            & ~ & 0.206  & &    2000 Oct 11  & ACIS-S & FAINT  & 35.8   &   10 17 03.8  & 39 02 49  \\
RXJ0439.0+0520       & ~ & 0.208  & &    2000 Aug 29  & ACIS-I & VFAINT & 7.6    &   04 39 02.3  & 05 20 44  \\
RXJ1504.1-0248       & ~ & 0.215  & &    2005 Mar 20  & ACIS-I & VFAINT & 29.4   &   15 04 07.9  &-02 48 16  \\
RXJ2129.6+0005       & ~ & 0.235  & &    2000 Oct 21  & ACIS-I & VFAINT & 7.6    &   21 29 39.9  & 00 05 20  \\
Abell 1835(1)        & ~ & 0.252  & &    1999 Dec 11  & ACIS-S & FAINT  & 18.0   &   14 01 01.9  & 02 52 43  \\
Abell 1835(2)        & ~ &  ``    & &    2000 Apr 29  & ACIS-S & FAINT  & 10.3   &      ``       &    ``     \\
Abell 611            & ~ & 0.288  & &    2001 Nov 03  & ACIS-S & VFAINT & 34.5   &   08 00 56.8  & 36 03 24  \\
Zwicky 3146          & ~ & 0.291  & &    2000 May 10  & ACIS-I &  FAINT & 41.4   &   10 23 39.4  & 04 11 14  \\
Abell 2537           & ~ & 0.295  & &    2004 Sep 09  & ACIS-S & VFAINT & 36.0   &   23 08 22.1  &-02 11 29  \\
MS2137.3-2353(1)     & ~ & 0.313  & &    1999 Nov 18  & ACIS-S & VFAINT & 20.5   &   21 40 15.2  &-23 39 40  \\
MS2137.3-2353(2)     & ~ &  ``    & &    2003 Nov 18  & ACIS-S & VFAINT & 26.6   &       ``      &    ``     \\
MACSJ0242.6-2132     & ~ & 0.314  & &    2002 Feb 07  & ACIS-I & VFAINT & 10.2   &   02 42 35.9  &-21 32 26  \\
MACSJ1427.6-2521     & ~ &        & &    2002 Jun 29  & ACIS-I & VFAINT & 14.7   &   14 27 39.4  &-25 21 02  \\
MACSJ2229.8-2756     & ~ & 0.324  & &    2002 Nov 13  & ACIS-I & VFAINT & 11.8   &   22 29 45.3  &-27 55 37  \\
MACSJ0947.2+7623     & ~ & 0.345  & &    2000 Oct 20  & ACIS-I & VFAINT & 9.6    &   09 47 13.1  & 76 23 14  \\
MACSJ1931.8-2635     & ~ & 0.352  & &    2002 Oct 20  & ACIS-I & VFAINT & 12.2   &   19 31 49.6  &-26 34 34  \\
MACSJ1115.8+0129     & ~ &        & &    2003 Jan 23  & ACIS-I & VFAINT & 10.2   &   11 15 52.1  & 01 29 53  \\
MACSJ1532.9+3021(1)  & ~ & 0.363  & &    2001 Aug 26  & ACIS-S & VFAINT & 9.4    &   15 32 53.9  & 30 20 59  \\
MACSJ1532.9+3021(2)  & ~ &  ``    & &    2001 Sep 06  & ACIS-I & VFAINT & 9.2    &       ``      &    ``     \\
MACSJ0011.7-1523(1)  & ~ &        & &    2002 Nov 20  & ACIS-I & VFAINT & 18.2   &   00 11 42.9  &-15 23 22  \\
MACSJ0011.7-1523(2)  & ~ &  ``    & &    2005 Jun 28  & ACIS-I & VFAINT & 32.1   &       ``      &    ``     \\
MACSJ1720.3+3536(1)  & ~ & 0.391  & &    2002 Nov 03  & ACIS-I & VFAINT & 16.6   &   17 20 16.8  & 35 36 27  \\
MACSJ1720.3+3536(2)  & ~ &  ``    & &    2005 Nov 22  & ACIS-I & VFAINT & 24.8   &       ``      &    ``     \\
MACSJ0429.6-0253     & ~ & 0.399  & &    2002 Feb 07  & ACIS-I & VFAINT & 19.1   &   04 29 36.1  &-02 53 08  \\
MACSJ0159.8-0849(1)  & ~ &        & &    2002 Oct 02  & ACIS-I & VFAINT & 14.1   &   01 59 49.4  &-08 49 58  \\
MACSJ0159.8-0849(2)  & ~ &  ``    & &    2004 Dec 04  & ACIS-I & VFAINT & 28.9   &       ``      &    ``     \\
MACSJ0329.7-0212(1)  & ~ & 0.450  & &    2002 Dec 24  & ACIS-I & VFAINT & 16.8   &   03 29 41.7  &-02 11 48  \\
MACSJ0329.7-0212(2)  & ~ &  ``    & &    2004 Dec 06  & ACIS-I & VFAINT & 31.1   &       ``      &    ``     \\
RXJ1347.5-1145(1)    & ~ & 0.451  & &    2000 Mar 03  & ACIS-S & VFAINT & 8.6    &   13 47 30.6  &-11 45 10  \\
RXJ1347.5-1145(2)    & ~ &  ``    & &    2000 Apr 29  & ACIS-S & FAINT  & 10.0   &       ``      &    ``     \\
RXJ1347.5-1145(3)    & ~ &  ``    & &    2003 Sep 03  & ACIS-I & VFAINT & 49.3   &       ``      &    ``     \\
3C295(1)             & ~ & 0.461  & &    1999 Aug 30  & ACIS-S & FAINT  & 15.4   &   14 11 20.5  & 52 12 10  \\
3C295(2)             & ~ &  ``    & &    2001 May 18  & ACIS-I & FAINT  & 72.4   &       ``      &    ``     \\
MACSJ1621.6+3810(1)  & ~ & 0.461  & &    2002 Oct 18  & ACIS-I & VFAINT & 7.9    &   16 21 24.8  & 38 10 09  \\
MACSJ1621.6+3810(2)  & ~ &  ``    & &    2004 Dec 11  & ACIS-I & VFAINT & 32.2   &       ``      &    ``     \\
MACSJ1621.6+3810(3)  & ~ &  ``    & &    2004 Dec 25  & ACIS-I & VFAINT & 26.1   &       ``      &    ``     \\
MACSJ1311.0-0311     & ~ & 0.494  & &    2005 Apr 20  & ACIS-I & VFAINT & 56.2   &   13 11 01.6  &-03 10 40  \\
MACSJ1423.8+2404     & ~ & 0.539  & &    2003 Aug 18  & ACIS-S & VFAINT & 113.5  &   14 23 47.9  & 24 04 43  \\
MACSJ0744.9+3927(1)  & ~ & 0.686  & &    2001 Nov 12  & ACIS-I & VFAINT & 17.1   &   07 44 52.9  & 39 27 27  \\
MACSJ0744.9+3927(2)  & ~ &  ``    & &    2003 Jan 04  & ACIS-I & VFAINT & 15.6   &       ``      &    ``     \\
MACSJ0744.9+3927(3)  & ~ &  ``    & &    2004 Dec 03  & ACIS-I & VFAINT & 41.3   &       ``      &    ``     \\
\hline                      
\end{tabular}
\end{center}
\end{table*}

\subsection{Target selection}\label{section:targets}

Our target clusters are the most massive, dynamically relaxed clusters
known in the redshift range $0<z<0.7$.  They form a restricted set of
the clusters used by Allen \etal (2007, in preparation) to study the
evolution of the X-ray gas mass fraction and constrain cosmological
parameters. In detail, we have used only the targets from that study
for which the temperature is measured in at least four independent
bins, which permits reliable measurements on the inner density
slopes. The target clusters all have mass weighted X-ray temperatures
measured within the radius $r_{2500}$\footnote{$r_{2500}$ is the
radius within which the mean mass density is 2500 times the critical
density of the Universe at the redshift of the cluster.},
$kT_{2500}\approxgt5$\,keV (Allen \etal 2007, in preparation).

The clusters exhibit a high degree of dynamical relaxation in their
Chandra images, with sharp central X-ray surface brightness peaks,
regular elliptical X-ray isophotes and minimal isophote centroid
variations.  The clusters show minimal evidence for departures from
hydrostatic equilibrium in X-ray pressure maps (Million \etal, in
preparation).  The exceptions to this are RXJ1347.5-1145, and
MACSJ0744.9+3927, for which clear substructure is observed between
position angles of 90-190 degrees and 210-330 degrees,
respectively. These regions, associated with obvious substructure,
have been excluded from the analysis. The restriction to clusters with
the highest possible degree of dynamical relaxation (for which the
assumption of hydrostatic equilibrium should be most valid) minimizes
systematic scatter and allows the most precise test of the CDM model
predictions \citep[e.g., ][]{Nagai06}.

\subsection{Observations, data reduction and spectral analysis}
\label{section:obs}

The Chandra observations were carried out using the Advanced CCD
Imaging Spectrometer (ACIS) between 1999 August 30 and 2005 June 28.
The standard level-1 event lists produced by the Chandra pipeline
processing were reprocessed using the $CIAO$ (version 3.2.2) software
package, including the latest gain maps and calibration products. Bad
pixels were removed and standard grade selections applied. Where
possible, the extra information available in VFAINT mode was used to
improve the rejection of cosmic ray events. The data were cleaned to
remove periods of anomalously high background using the standard 
energy ranges and time bins recommended by the Chandra X-ray Center. 
The net exposure times after cleaning are summarized in 
Table~\ref{table:obs}. 

The data have been analysed using techniques discussed by \citet{Allen04}
and references therein.  In brief, concentric annular spectra
were extracted from the cleaned event lists, centred on the
coordinates listed in Table~\ref{table:obs}. Emission associated with
X-ray point sources or obvious substructure
(Section~\ref{section:targets}) was excluded.  The spectra were
analysed using XSPEC (version 11.3: \citealt{Arnaud96}), the MEKAL plasma
emission code (\citealt{Kaastra93}; incorporating the Fe-L
calculations of \citealt{Liedahl95}) and the
photoelectric absorption models of \citet{Balucinska92}.
We have included standard correction factors to account for
time-dependent contamination along the instrument light path.  In
addition, we have incorporated a small correction to the High
Resolution Mirror Assembly model in CIAO 3.2.2, which takes the form
of an 'inverse' edge with an energy, E=2.08\,keV and optical depth
$\tau=-0.1$ (H. Marshall, private communication) and boosted the
overall effective area by six per cent, to better match later
calibration data (A. Vikhlinin, private communication). Only data in
the $0.8-7.0$ keV energy range were used in the analysis (the
exceptions being the earliest observations of 3C 295, Abell 1835 and
Abell 2029 where a wider 0.6 to 7.0 keV band was used).

For the nearer clusters ($z<0.3$), background spectra were extracted
from the blank-field data sets available from the Chandra X-ray
Center. These were cleaned in an identical manner to the target
observations. In each case, the normalizations of the background files
were scaled to match the count rates in the target observations
measured in the 9.5-12keV band. Where required, \eg due to the
presence of strong excess soft emission in the field of Abell 2029, a
spectral model for any unusual soft background emission was included
in the analysis. For the more distant systems (as well as for the
first observation of Abell 1835, the ACIS-I observation of Abell 383,
and the observations of Abell 2537, RXJ 2129.6+0005 and Zwicky 3146)
background spectra were extracted from appropriate, source free
regions of the target data sets. (We have confirmed that similar
results are obtained using the blank-field background data sets.) In
order to minimize systematic uncertainties and due to the specific
goals of this work, we have restricted our spectral analysis to radii
within which systematic uncertainties in the background subtraction
(established by the comparison of different background subtraction
methods) are smaller than the statistical uncertainties in the
results. All results are drawn from spectral analyses limited to ACIS
chips 0,1,2,3 and 7 which have the most accurate calibration, although
ACIS chip 5 was also used to study the soft X-ray background in ACIS-S
observations. We do not attempt to extend our analyses to larger radii
using the data from other chips.

Separate photon-weighted response matrices and effective area files
were constructed for each region using calibration files appropriate
for the period of observations. The spectra for all annuli for
a given cluster were modelled simultaneously in order to determine the
deprojected X-ray gas temperature and metallicity profiles, under the
assumption of spherical symmetry.

\subsection{Cluster mass profile measurements}
\label{fitting}

\subsubsection{Basics of the mass analysis}

Under the assumptions of hydrostatic equilibrium and spherical
symmetry, the observed X-ray surface brightness profile and the
deprojected X-ray gas temperature profile may together be used to
determine the X-ray emitting gas mass and total mass profile of a
galaxy cluster. For this analysis, we have used an enhanced version of
the Cambridge X-ray deprojection code described by \eg
\citet*{White97}.  This method is particularly well suited to the
present study in that it does not require approximate fitting
functions for the X-ray temperature, gas density or surface brightness
when measuring the total, gravitating mass\footnote{As discussed in the
text, priors are required when modelling the dark matter and X-ray and
optically luminous matter components separately}. The use of such functions
introduces priors into an analysis which can complicate the
interpretation of results and, in particular, the estimation of
measurement errors.

We have carried out two separate mass analyses: firstly (method 1) an 
analysis in which the total mass profile (dark plus luminous
matter) was modelled using either an NFW or singular isothermal sphere.
Detailed results on the X-ray emitting gas profiles were also determined 
at this stage. Secondly (method 2) an analysis
in which the total mass profile was modelled as the sum of 
three parts: the dark
matter halo (fitted with a generalized NFW profile), the X-ray
emitting gaseous halo (approximated, for the purposes of this analysis only, 
with a beta model), and the optically
luminous mass of the cD galaxy (approximated with a Jaffe or de Vaucouleurs
model).

The normalization, $\rho_0$, of the generalized NFW mass model
(eq.~\ref{gnfw}) is usually written as

\begin{equation}
\rho_0=\rho_{\rm crit}\, \delta_c
\end{equation}
where
\begin{equation}
\rho_{\rm crit} = 3H(z)^2/ 8 \pi G
\end{equation}
is the critical density at the redshift $z$ of the cluster. $G$ is
the gravitational constant and the Hubble parameter $H(z)$ in the flat
$\Lambda$CDM reference cosmology is defined by
\begin{equation}
H(z)^2=H_0^2\, \left((1+z)^3\Omega_{\rm m}+1-\Omega_{\rm m}\right),
\end{equation}
where $\Omega_{\rm m}$ is the matter density in units of the critical
density.

The amplitude $\delta_c$ depends only on the concentration parameter
$c=\frac{r_{\rm s}}{r_{200}}$ ($r_{200}$ is the radius within which
the average mass density is 200 times $\rho_{\rm crit}$). For convenient
computation, $\delta_c$ can be written using the Gauss hypergeometric
function F(a,b;c;z) \citep{Abramowitz65}
\begin{equation}
\Phi(y)=\frac{y^{3-\alpha}}{3-\alpha} F(3-\alpha,3-\alpha;4-\alpha;-y)
\end{equation}
as
\begin{equation}
\delta_{\rm c} = \frac{200}{3} \frac{ c^3}{\Phi(c)}
\end{equation}
\citep{Wyithe01}. For the original NFW mass model with $\alpha=1$, one
obtains $\Phi(c)={\rm ln}(1+c)-c/(1+c)$ \citep{Navarro96}.

With mass analysis method 2, the component of the 
total mass distribution representing the X-ray emitting gas 
component was described by a beta-model \citep{Cavaliere78}
\begin{equation}
\rho_{\rm gas}(r)=\rho_{0,{\rm gas}} \left[1+\left(\frac{r}{r_{\rm
      c,\,gas}}\right)^2\right]^{-\frac{3\beta}{2}},
\end{equation}
where $\rho_{0,{\rm gas}}$ is the central gas density, $r_{\rm
  c,\,gas}$ is the core radius of the gas profile and $\beta$ is the
slope parameter. (We stress that the detailed results on the
X-ray emitting gas profiles, used for example in the 
measurement of cluster gas mass fractions, are determined with method 1
and involve no assumption about the functional form of the gas profile.
The use of the beta-model with method 2 simply approximates 
the contribution of this mass component to the total mass, 
allowing us to recover the dark matter profiles. Although the 
beta model does not provide a precise match to the X-ray gas mass 
distribution in all cases, the systematic uncertainties in the dark 
matter profiles that result from its 
use are small, primarily because the X-ray emitting gas contributes 
only $\sim 12$ per cent of the total mass; Allen \etal 2007, in preparation).

All of the target clusters have a single, optically dominant galaxy at
their centres. With method 2, we accounted for the mass of stars in
the central galaxy using a
\citet{Jaffe83} model \citep{Sand02,Sand04}. This was added to the generalized
NFW potential for the dark matter halo and the beta-model for the
X-ray emitting gas to obtain the total mass profile.
The \citet{Jaffe83} model is 
\begin{equation}
\rho_{\rm J}(r)=\frac{\rho_{0,{\rm J}}}{(\frac{r}{r_{\rm c}})^2 (1+
\frac{r}{r_{\rm c}})^2}
\end{equation}
with central density $\rho_{0,{\rm J}}$ and core radius $r_{\rm c}$.
The total mass for the model is finite, $M_{\rm J}=4\pi\,r_{\rm
c}^3\,\rho_{0,{\rm J}}$. Since including this model into the analysis
has only a small effect on the results (Section~\ref{stars}) but
requires two parameters, we chose to fix these two parameters to
sensible values. We adopt $R_{\rm e}=0.76\,r_{\rm c}=25$ kpc
as a typical effective radius
\citep[e.g., ][]{Sand02,Sand05}.  For the well-studied cD galaxy in
MS2137-2353, \citet{Sand02} measure a total V-band luminosity of
$L_V=4.16\times10^{11}\,L_{\odot}$.  Using the $M/L_V$
relation of \citet{Fukugita98}
\begin{equation}
\frac{M}{L_V}=4.0+0.38\, (t_{\rm g} - 10\,{\rm Gyr}),
\end{equation}
where $t_{\rm g}$ is the age of the galaxy (we assume a formation
redshift $z_{\rm f}=2.0$), we estimate the total stellar mass
associated with the cD galaxy of $1.14\times10^{12}$\Msun. The central
galaxy for each cluster in the sample is assumed to have this
stellar mass.

We have also carried out a repeat analysis in which the
\citet{Jaffe83} model for the stellar mass associated with the
central, dominant galaxies was replaced by a \citet{deVaucouleurs48}
model. This analytical model is described by the surface mass density
\begin{equation}
\Sigma(r)=\Sigma_{\rm e}\, {\rm e}^{-7.67 \left[(r/R_{\rm
e})^{1/4}-1\right]},
\end{equation}
where $\Sigma_{\rm e}$ is the surface density at the effective (or
half-mass) radius $R_{\rm e}$. The 3-dimensional mass profile of the
de Vaucouleurs model was calculated by \citet{Young76} and has a
shallower central slope close to unity. However, similar results on
the cluster dark matter profiles were obtained in all cases, showing
that the precise choice of the galaxy model has a negligible effect on the
results.

\subsubsection{Best-fitting values and confidence limits}

Given the observed surface brightness profile and a particular
parameterized model for the total mass, the deprojection code is 
used to predict the
temperature profile of the X-ray gas. This model temperature profile
is compared with the observed spectral, deprojected temperature
profile and the goodness of fit is calculated using the sum over all
temperature bins

\begin{equation} \chi^2 = \sum_{\,\rm all\,bins}\,\left(
\frac{T_{\,\rm obs} - T_{\,\rm model}}{\sigma_{\,\rm obs}} \right)^2,
\end{equation} 
where $T_{\,\rm obs}$ is the observed, spectral deprojected
temperature profile and $T_{\,\rm model}$ is the model, rebinned to
the same spatial scale using flux weighting.\footnote{In detail, we
use the median model temperature profile determined from 100
Monte-Carlo simulations. The outermost pressure, at the limit of the
X-ray surface brightness profile, is fixed using an iterative method
that ensures a smooth, power law pressure gradient in these regions. The model
temperature profiles, for radii spanned by the spectral data, are not
sensitive to reasonable choices for the outer pressures.}

For each mass model, we determine the best-fitting parameter values
and uncertainties via $\chi^2$ minimization. We use the LEASQR
Levenberg-Marquardt routine \citep{Marquardt63} from the GNU Octave
Repository, available online at http://sourceforge.net. (A full Monte
Carlo analysis, as described above, is run for each set of parameter
values examined in the minimization procedure.)

\begin{table}
\begin{center}
\caption{Regions with residual substructure that were
downweighted in the mass analysis. A systematic uncertainty of 
$\pm30$ per cent has been added in quadrature to all 
temperature measurements made within radii $R_{\rm sub}$.}\label{table:exclude}
\begin{tabular}{ l c c }
&&  \\ 
                    & ~ &    $R_{\rm sub}$ (kpc)\\
\hline
Abell 1795          & ~ &    73.3  \\
Abell 2029          & ~ &    30.0  \\
Abell 478           & ~ &    14.6  \\
PKS0745-191         & ~ &    53.0  \\
Abell 1413          & ~ &    38    \\
Abell 2204          & ~ &    76.8  \\
Abell 383           & ~ &    38.7  \\
RXJ1504.1-0248      & ~ &    79.1  \\
RXJ2129.6+0005      & ~ &    40.4  \\
Zwicky 3146         & ~ &    242   \\
Abell 2537          & ~ &    41.1  \\
MACSJ2229.8-2756    & ~ &    42.0  \\
MACSJ0947.2+7623    & ~ &    40.1  \\
MACSJ1931.8-2635    & ~ &    41.5  \\
MACSJ1115.8+0129    & ~ &    83.4  \\
MACSJ1532.9+3021    & ~ &    42.3  \\
MACSJ1621.6+3810    & ~ &    43.1  \\
\hline                      
\end{tabular}
\end{center}
\end{table}

For a number of the clusters, the Chandra images indicate small levels
of residual substructure in the innermost regions, which probably
result from `sloshing' of the X-ray emitting gas within the central
potentials \citep[e.g., ][]{Markevitch01,Ascasibar06,Allen92}
and/or interactions between
central radio sources and the surrounding intracluster gas 
(e.g., \citealt{Boehringer93,Fabian00,Fabian03,Fabian06,Birzan04,Dunn04,Forman05,Dunn05,Rafferty06};
Allen et al. 2006).
The regions affected by such substructure are
listed in Table~\ref{table:exclude}. A systematic uncertainty of $\pm
30$ per cent has been added in quadrature to the spectral results from
these regions which leads to them having little weight in the mass
analysis.

\section{results}

\subsection{Total mass profiles: NFW versus singular isothermal sphere}
\label{section:cm}

\begin{table*}
\begin{center}
\caption{Results from fits to the 
total mass profiles with NFW (slope $\alpha=1$ fixed) and singular-isothermal (SI) 
models (analysis method 1).
Columns list the number of temperature bins used in the 
analysis of each cluster, the scale radius $r_{\rm s}$ (in kpc) 
and concentration parameter $c=r_{200}/r_{\rm s}$ for the NFW models, 
and the goodness of fit ($\chi^2$/DOF) for both models. The SI model
provides a poor description of the data. Quoted uncertainties are
68 per cent ($\Delta \chi^2=1.0$) confidence limits.}
\label{table:totalmass}
\vskip -0.1truein
\begin{tabular}{ l c c c c c c c c }
&&&&&&  \\
& & \multicolumn{7}{c}{TOTAL MASS (LUMINOUS PLUS DARK MATTER)} \\
& & Number of & & \multicolumn{3}{c}{ NFW MODEL} & & SI MODEL\\
& & $kT$ bins & & $r_{\rm s}$ & $c$ & $\chi^2$/DOF & & $\chi^2$/DOF\\
\hline                                                                      
Abell 1795             & & 5 &  & $0.43^{+0.10}_{-0.10}$ & $4.59^{+0.86}_{-0.62}$& 2.06/3 & & 49.2/4  \\
Abell 2029             & & 6 &  & $0.28^{+0.03}_{-0.02}$ & $6.95^{+0.30}_{-0.39}$& 4.73/4 & & 303/5   \\
Abell 478              & & 7 &  & $0.50^{+0.05}_{-0.06}$ & $4.48^{+0.40}_{-0.25}$& 8.85/5 & & 615/6   \\
PKS0745-191            & & 7 &  & $0.34^{+0.13}_{-0.08}$ & $6.43^{+1.22}_{-1.24}$& 5.13/5 & & 42.1/6  \\
Abell 1413             & & 6 &  & $0.48^{+0.13}_{-0.09}$ & $4.32^{+0.69}_{-0.64}$& 24.6/5 & & 91.5/6  \\
Abell 2204             & & 5 &  & $0.19^{+0.08}_{-0.05}$ & $9.84^{+2.34}_{-2.13}$& 4.38/3 & & 16.6/4  \\
Abell 383              & & 5 &  & $0.48^{+0.12}_{-0.12}$ & $3.75^{+0.71}_{-0.48}$& 23.3/4 & & 90.0/5  \\
Abell 963              & & 5 &  & $0.39^{+0.12}_{-0.08}$ & $4.73^{+0.84}_{-0.77}$& 6.16/3 & & 97.8/4  \\
RXJ0439.0+0521         & & 4 &  & $0.19^{+0.06}_{-0.04}$ & $7.69^{+1.47}_{-1.24}$& 0.30/2 & & 28.3/3  \\
RXJ1504.1-0248         & & 5 &  & $0.54^{+0.29}_{-0.15}$ & $4.38^{+1.06}_{-1.04}$& 0.91/3 & & 28.1/4  \\
RXJ2129.6+0005         & & 5 &  & $0.38^{+0.45}_{-0.17}$ & $4.59^{+2.18}_{-1.84}$& 1.53/3 & & 14.7/4  \\
Abell 1835             & & 5 &  & $0.58^{+0.08}_{-0.09}$ & $4.20^{+0.43}_{-0.30}$& 12.5/3 & & 302/4   \\
Abell 611              & & 5 &  & $0.32^{+0.20}_{-0.10}$ & $5.39^{+1.60}_{-1.51}$& 2.35/3 & & 25.7/4  \\
Zwicky 3146            & & 5 &  & $0.99_{-0.57}$         & $2.71^{+2.23}_{-2.71}$& 0.79/3 & & 11.4/4  \\
Abell 2537             & & 5 &  & $0.37^{+0.31}_{-0.15}$ & $4.86^{+2.06}_{-1.62}$& 3.42/3 & & 22.2/4  \\
MS2137.3-2353          & & 6 &  & $0.18^{+0.02}_{-0.02}$ & $8.19^{+0.54}_{-0.56}$& 5.52/4 & & 235/5   \\
MACSJ0242.6-2132       & & 4 &  & $0.19^{+0.05}_{-0.03}$ & $7.89^{+1.07}_{-1.04}$& 2.92/2 & & 62.1/3  \\
MACSJ1427.6-2521       & & 4 &  & $0.15^{+0.08}_{-0.05}$ & $8.18^{+2.28}_{-1.92}$& 4.77/2 & & 18.1/3  \\
MACSJ2229.8-2756       & & 4 &  & $0.15^{+0.10}_{-0.05}$ & $8.43^{+3.27}_{-2.53}$& 1.69/2 & & 6.8/3   \\
MACSJ0947.2+7623       & & 4 &  & $0.32^{+0.18}_{-0.11}$ & $6.01^{+1.77}_{-1.44}$& 2.53/2 & & 23.6/3  \\
MACSJ1931.8-2635       & & 4 &  & $0.51^{+0.95}_{-0.20}$ & $4.05^{+1.54}_{-1.91}$& 1.98/2 & & 33.6/3  \\
MACSJ1115.8+0129       & & 4 &  & $1.35_{-0.90}$         & $2.15^{+2.19}_{-2.15}$& 1.23/2 & & 13.7/3  \\
MACSJ1532.9+3021       & & 5 &  & $0.34^{+0.16}_{-0.09}$ & $5.29^{+1.22}_{-1.14}$& 1.76/3 & & 41.8/4  \\
MACSJ0011.7-1523       & & 4 &  & $0.49^{+0.17}_{-0.13}$ & $3.75^{+0.79}_{-0.63}$& 6.75/2 & & 168/3   \\
MACSJ1720.3+3536       & & 4 &  & $0.33^{+0.13}_{-0.09}$ & $5.21^{+1.06}_{-0.98}$& 1.19/2 & & 49.8/3  \\
MACSJ0429.6-0253       & & 4 &  & $0.16^{+0.04}_{-0.03}$ & $8.50^{+1.40}_{-1.18}$& 1.84/2 & & 25.8/3  \\
MACSJ0159.8-0849       & & 4 &  & $0.37^{+0.10}_{-0.10}$ & $5.28^{+1.19}_{-0.74}$& 4.41/2 & & 52.9/3  \\
MACSJ0329.7-0212       & & 4 &  & $0.29^{+0.07}_{-0.05}$ & $5.48^{+0.76}_{-0.67}$& 6.24/2 & & 86.0/3  \\
RXJ1347.5-1144         & & 6 &  & $0.45^{+0.10}_{-0.07}$ & $5.63^{+0.54}_{-0.60}$& 16.1/4 & & 257/5   \\
3C295                  & & 4 &  & $0.15^{+0.03}_{-0.02}$ & $8.63^{+0.99}_{-0.89}$& 1.12/2 & & 61.3/3  \\
MACSJ1621.6+3810       & & 4 &  & $0.25^{+0.17}_{-0.10}$ & $6.37^{+2.78}_{-1.91}$& 4.00/2 & & 17.6/3  \\
MACSJ1311.0-0311       & & 4 &  & $0.31^{+0.15}_{-0.09}$ & $4.91^{+1.21}_{-1.10}$& 1.54/2 & & 25.4/3  \\
MACSJ1423.8+2404       & & 4 &  & $0.17^{+0.03}_{-0.02}$ & $8.27^{+0.72}_{-0.68}$& 0.05/2 & & 74.1/3  \\
MACSJ0744.9+3927       & & 4 &  & $0.32^{+0.18}_{-0.09}$ & $4.88^{+1.13}_{-1.23}$& 4.40/2 & & 40.9/3  \\
\hline                                                                                                                                       
\end{tabular}                                                                                                                                
\end{center}
\end{table*}

Table~\ref{table:totalmass} summarizes the results from the initial
mass analysis (method 1) in which the total mass profiles
(dark-plus-luminous matter) were modelled using either an NFW (inner
slope $\alpha=1$ fixed) or singular isothermal sphere model ($\rho(r)
= A/r^2$, with the normalization $A$ free). We see that for most
clusters the NFW model provides a reasonable description of the total
mass profiles. For 27 out of 34 clusters, the
probability of the $\chi^2$/DOF value based on a $\chi^2$ distribution
is $0.05$ or better \citep[e.g.,][]{Bevington92}
and only for Abell 383 and 1413 does the
probability drop below $0.001$.  In contrast, the singular isothermal
sphere model can be firmly rejected for most clusters in the sample.

Combining the results for all clusters, the NFW model gives a total
$\chi^2$ of 171.0 for 95 degrees of freedom (DOF) (with Abell 383
and 1413 contributing a total $\chi^2$ of 47.9). 
The singular isothermal sphere gives $\chi^2=3031.1$ for 129 DOF, 
indicating an extremely low model probability.

\subsection{Dark matter profiles: The NFW model and mass-concentration relation}
\label{dark}

\begin{table*}
\caption{Results on the dark matter profiles for the clusters, 
from the fits including separate components for the 
X-ray emitting gas and dominant cluster galaxy (analysis method 2). 
For the standard NFW model (inner slope $\alpha=1$ fixed) 
we list the scale radius $r_{\rm s}$ (in kpc) 
and concentration parameter $c=r_{200}/r_{\rm s}$. 
For the generalized NFW model, we give results on the 
inner slope $\alpha$. 
For those cases where the formal best-fitting values for the
slope are negative, we only list the 95 per cent upper limit. Where
confidence limits are absent in the table, the fit results were
consistent with $\alpha=0$. For both models, we list the results on
the goodness of fit ($\chi^2$/DOF). Quoted uncertainties are 68 per
cent ($\Delta \chi^2=1.0$) confidence limits.}
\label{table:dm}
\begin{center}
\begin{tabular}{ l c c c c c l c}
&&&&&&&  \\
&{~~} & \multicolumn{6}{c}{DARK MATTER PROFILES} \\
& {~~} & \multicolumn{3}{c}{NFW ($\alpha=1$)} & {~~} &
\multicolumn{2}{c}{GENERALIZED NFW} \\
                       & &    $r_{\rm s}$           &         $c$            & $\chi^2$/DOF  & &   $\alpha$ &     $\chi^2$/DOF \\ 
\hline                                                                      
Abell 1795             & &   $0.41^{+0.13}_{-0.09}$ & $4.45^{+0.85}_{-0.77}$ &  2.21/3  &~& $      <1.63         $    & 1.57/2 \\
Abell 2029             & &   $0.28^{+0.03}_{-0.02}$ & $6.63^{+0.34}_{-0.37}$ &  4.47/4  & & $1.16^{+0.22}_{-0.34}$    & 4.31/3 \\
Abell 478              & &   $0.56^{+0.09}_{-0.08}$ & $3.91^{+0.36}_{-0.33}$ &  6.90/5  & & $1.10^{+0.12}_{-0.21}$    & 6.66/4 \\
PKS0745-191            & &   $0.36^{+0.13}_{-0.11}$ & $5.85^{+1.55}_{-1.07}$ &  5.26/5  & & $0.78^{+0.64}_{     }$    & 5.17/4 \\
Abell 1413             & &   $0.43^{+0.14}_{-0.09}$ & $4.43^{+0.78}_{-0.75}$ &  24.5/5  & & $1.54^{+0.06}_{-0.21}$    & 22.0/4 \\
Abell 2204             & &   $0.18^{+0.08}_{-0.05}$ & $9.75^{+2.92}_{-2.17}$ &  4.93/3  & & $      <1.50         $    & 3.14/2 \\
Abell 383              & &   $0.45^{+0.16}_{-0.08}$ & $3.76^{+0.53}_{-0.68}$ &  23.5/4  & & $      <0.80         $    & 18.7/2 \\
Abell 963              & &   $0.40^{+0.16}_{-0.10}$ & $4.38^{+0.88}_{-0.88}$ &  7.21/3  & & $      <1.02         $    & 3.73/2 \\
RXJ0439.0+0521         & &   $0.21^{+0.08}_{-0.06}$ & $6.65^{+1.54}_{-1.21}$ &  0.62/2  & & $      <1.59         $    & 0.01/1 \\
RXJ1504.1-0248         & &   $0.62^{+0.43}_{-0.19}$ & $3.76^{+1.05}_{-1.09}$ &  0.88/3  & & $0.92^{+0.61}_{     }$    & 0.87/2 \\
RXJ2129.6+0005         & &   $0.41^{+0.72}_{-0.20}$ & $4.06^{+2.31}_{-1.97}$ &  1.51/3  & & $1.08^{+0.52}_{     }$    & 1.49/2 \\
Abell 1835             & &   $0.71^{+0.12}_{-0.14}$ & $3.42^{+0.45}_{-0.31}$ &  14.2/3  & & $0.14^{+0.63}_{     }$    & 11.3/2 \\
Abell 611              & &   $0.32^{+0.24}_{-0.11}$ & $5.08^{+1.72}_{-1.62}$ &  2.43/3  & & $0.64^{+0.94}_{     }$    & 2.37/2 \\
Zwicky 3146            & &   $1.14^{+>5.0}_{-0.71}$ & $2.31^{+2.31}_{-2.31}$ &  0.82/3  & & $      <1.78         $    & 0.59/2 \\
Abell 2537             & &   $0.35^{+0.28}_{-0.15}$ & $4.82^{+2.32}_{-1.59}$ &  2.74/3  & & $      <1.66         $    & 1.51/2 \\
MS2137.3-2353          & &   $0.20^{+0.03}_{-0.02}$ & $7.21^{+0.57}_{-0.59}$ &  5.10/4  & & $1.00^{+0.25}_{-0.35}$    & 5.10/3 \\
MACSJ0242.6-2132       & &   $0.22^{+0.06}_{-0.05}$ & $6.68^{+1.23}_{-0.92}$ &  3.01/2  & & $0.66^{+0.65}_{     }$    & 2.88/1 \\
MACSJ1427.6-2521       & &   $0.17^{+0.12}_{-0.06}$ & $7.14^{+2.29}_{-2.05}$ &  4.92/2  & & $      <1.74         $    & 4.57/1 \\
MACSJ2229.8-2756       & &   $0.16^{+0.13}_{-0.07}$ & $7.70^{+3.66}_{-2.62}$ &  1.50/2  & & $1.68^{+0.15}_{     }$    & 1.02/1 \\
MACSJ0947.2+7623       & &   $0.35^{+0.24}_{-0.13}$ & $5.40^{+1.86}_{-1.51}$ &  2.66/2  & & $      <1.65         $    & 1.94/1 \\
MACSJ1931.8-2635       & &   $0.69^{+2.46}_{-0.36}$ & $3.11^{+1.87}_{-1.88}$ &  1.88/2  & & $1.16^{+0.36}_{     }$    & 1.86/1 \\
MACSJ1115.8+0129       & &   $1.61^{+>5.0}_{-1.14}$ & $1.80^{+2.17}_{-1.80}$ &  1.30/2  & & $      <1.65         $    & 0.97/1 \\
MACSJ1532.9+3021       & &   $0.37^{+0.23}_{-0.12}$ & $4.70^{+1.32}_{-1.24}$ &  2.01/3  & & $      <1.57         $    & 1.28/2 \\
MACSJ0011.7-1523       & &   $0.59^{+0.25}_{-0.19}$ & $3.11^{+0.84}_{-0.62}$ &  8.34/2  & & $      <0.84         $    & 3.33/1 \\
MACSJ1720.3+3536       & &   $0.40^{+0.18}_{-0.13}$ & $4.36^{+1.21}_{-0.88}$ &  1.12/2  & & $1.08^{+0.38}_{     }$    & 1.09/1 \\
MACSJ0429.6-0253       & &   $0.17^{+0.05}_{-0.04}$ & $7.64^{+1.57}_{-1.10}$ &  1.41/2  & & $1.44^{+0.23}_{-0.66}$    & 0.79/1 \\
MACSJ0159.8-0849       & &   $0.38^{+0.18}_{-0.10}$ & $4.93^{+1.01}_{-1.06}$ &  3.65/2  & & $1.44^{+0.07}_{-0.22}$    & 1.60/1 \\
MACSJ0329.7-0212       & &   $0.32^{+0.11}_{-0.07}$ & $4.74^{+0.75}_{-0.78}$ &  6.80/2  & & $      <1.38         $    & 5.07/1 \\
RXJ1347.5-1144         & &   $0.54^{+0.08}_{-0.11}$ & $4.79^{+0.68}_{-0.38}$ &  18.0/4  & & $      <0.75         $    & 9.84/3 \\
3C295                  & &   $0.16^{+0.03}_{-0.03}$ & $7.78^{+1.03}_{-0.90}$ &  1.71/2  & & $      <1.44         $    & 0.85/1 \\
MACSJ1621.6+3810       & &   $0.26^{+0.21}_{-0.11}$ & $5.97^{+2.95}_{-1.95}$ &  4.12/2  & & $      <1.78         $    & 3.90/1 \\
MACSJ1311.0-0311       & &   $0.33^{+0.18}_{-0.11}$ & $4.42^{+1.39}_{-1.06}$ &  1.46/2  & & $1.46^{+0.13}_{-1.08}$    & 0.94/1 \\
MACSJ1423.8+2404       & &   $0.18^{+0.03}_{-0.02}$ & $7.68^{+0.71}_{-0.79}$ &  0.19/2  & & $0.66^{+0.62}_{     }$    & 0.02/1 \\
MACSJ0744.9+3927       & &   $0.36^{+0.19}_{-0.13}$ & $4.31^{+1.43}_{-1.06}$ &  4.72/2  & & $      <1.42         $    & 2.71/1 \\
\hline
\end{tabular}
\end{center}
\end{table*}

Table~\ref{table:dm} summarizes the results from the analysis with
method 2, in which the cluster mass distributions were separated into
three parts: the dark matter halo (fitted with a generalized NFW
profile), the X-ray emitting gaseous halo (approximated with a beta model),
and the optically luminous mass of the central dominant galaxy (approximated
with a Jaffe or a de Vaucouleurs model).  In the first case, we examined
models in which the inner slope of the dark matter profile was fixed
at $\alpha=1$ i.e. the standard NFW model.

Our first conclusion is that, as with the analysis of the total mass
profiles, the NFW model provides a good overall description of the
dark matter profiles in the clusters.  For 27 out of 34 clusters,
the $\chi^2$/DOF value has a probability of $0.05$ or better
and only for Abell 383 and 1413 does the model fail
significantly. Combining the results for all clusters, the NFW model
gives a total $\chi^2$ of  176.1 for 95 degrees of freedom (DOF) (with
Abell 383 and 1413 contributing a total $\chi^2$ of 48.0; note that the 
inclusion of the separate mass components for the
X-ray emitting gas and stars introduces no additional free parameters
in the fits.) The concentration parameters, $c$ for the dark matter
profiles are slightly lower, and the scale radii slightly higher, 
than for the total matter distributions. However, in general the 
results are quite similar.

\begin{figure}
\begin{center}
\resizebox{\columnwidth}{!}{\includegraphics{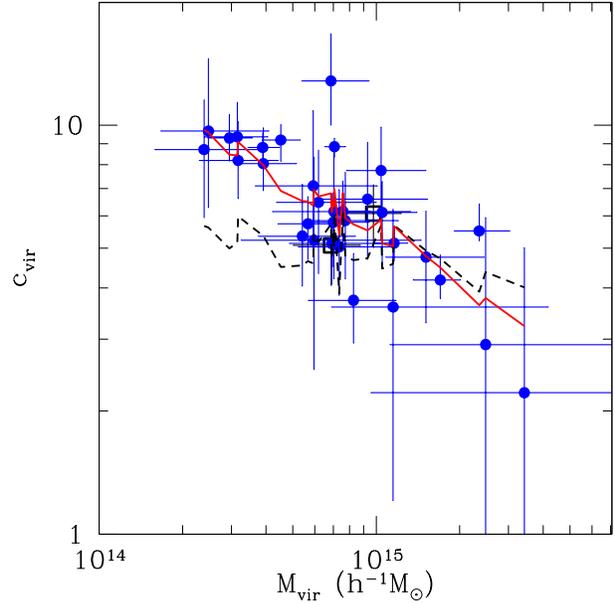}}
\end{center}
\caption{The virial mass-concentration relation for the dark
matter halos. The solid line shows the best-fitting model with
$c_0=7.55\pm0.90$, $a=-0.45\pm0.12$ and $b=0.71\pm0.52$ (95 per cent
confidence limits). The dashed line shows the prediction from the CDM
simulations of \citet{Shaw05}, 
with $c_0=6.47$, $a=-0.12$ and $b=1.0$
fixed. The two clusters for which the NFW model fails to provide a
reasonable description of the Chandra data, Abell 383 and 1413, are
plotted with open square symbols and have been excluded from the
fits.}
\label{mvirc}
\end{figure}

\begin{figure}
\begin{center}
\resizebox{\columnwidth}{!}{\includegraphics{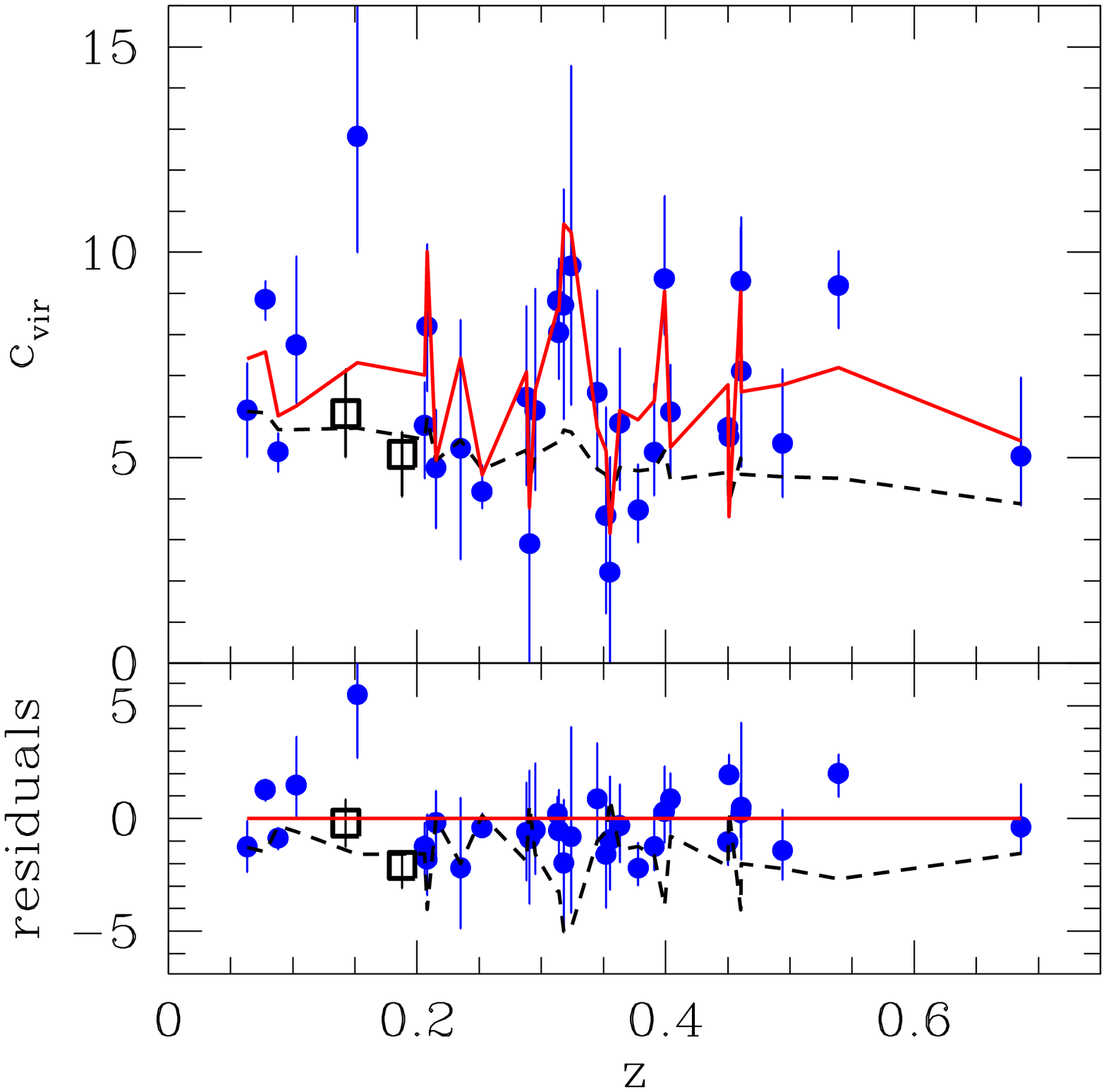}}
\end{center}
\caption{The measured concentration parameters for the dark matter halos 
as a function of redshift. The solid line shows the best-fitting model with
$c_0=7.55\pm0.90$, $a=-0.45\pm0.12$ and $b=0.71\pm0.52$ (95 per cent
confidence limits). The dashed line shows the prediction from the CDM
simulations of \citet{Shaw05} with $c_0=6.47$, $a=-0.12$ and $b=1.0$. 
The apparent absence of redshift evolution in the data may be an artifact 
induced by the steep slope of the underlying mass-concentration relation 
(see text for details). Symbols as in Fig.~\ref{mvirc}} 
\label{cz}
\end{figure}

As discussed in Section~\ref{section:intro}, one of the central
predictions from simulations of CDM halos is the mass-concentration
relation.  In order to allow for the most direct comparison with
theory, we have used the values in Tables~\ref{table:totalmass}
and~\ref{table:dm} to calculate the concentration parameters
$c_{\rm vir}$
and virial masses $M_{\rm vir}$ measured within the virial
radii for the clusters.
We adopt the definitions of virial radius and
virial mass used by \citet{Shaw05}
\begin{equation}
M_{\rm vir}=\frac{4}{3}\pi\,r_{\rm vir}^3\,\Delta_{\rm
  c}(z)\,\rho_{\rm crit}(z),
\end{equation}
where the virial overdensity is given by $\Delta_{\rm
c}=178\Omega_{\rm m}(z)^{0.45}$
\citep{Lahav91}.

Both the virial radii and virial masses are calculated for the
{\em total} mass model, including all mass components.
The concentration parameters are defined
by the ratios of these virial radii and the scale radii
of the dark matter components,
$c_{\rm vir}=r^{\rm total}_{\rm vir}/r^{\rm dark}_{\rm s}$.
Table~\ref{m-c} lists
the results on $c_{\rm vir}$ and $M_{\rm vir}$ for the clusters
using the standard NFW model ($\alpha=1$).

Fig.~\ref{mvirc} shows the variation of $c_{\rm vir}$ with $M_{\rm
vir}$ measured from the Chandra data. A tight relation is observed with a clear
trend for decreasing concentration parameter with increasing mass.

The literature contains a variety of simple analytical forms to
describe the expected form of the mass-concentration relation, based
on CDM numerical simulations. The simplest of these is a power-law
model \citep{Dolag04}

\begin{equation}
c_{\rm vir}(z)=\frac{c_0}{1+z} \left(\frac{M_{\rm vir}}{8\times10^{14}\,
      h^{-1}M_{\odot}}\right)^{a}.
\label{plaw}
\end{equation}
\citet{Shaw05} present one of the largest statistical studies of cluster-sized
CDM halos to date. Fitting a $z=0.05$ snapshot of their simulated data  
with the \citet{Dolag04} power law model, these authors find 
$c_0=6.47\pm0.03$ and $a=-0.12\pm0.03$ (68 per cent confidence limits). 
The dashed line in Fig.~\ref{mvirc} shows the model defined by these 
parameters, overlaid on the Chandra results. The model clearly provides 
a poor fit to the
observations, $\chi^2=125.9$ for
30 DOF, lying systematically below the data at lower masses and above 
it in the highest mass range. Note that the two clusters for which 
the NFW model provides a formally unacceptable description of the 
Chandra data, Abell 383 and 1413, have been
excluded from the fit and are plotted with open square symbols.
Note also that the mass-concentration relation of \citet{Shaw05}
is in good agreement with the models of \citet{Bullock01} (with the
parameters from \citealt{Dolag04}) and \citet{Eke01}.

One should remember that the Chandra observations reported here are
for the most massive, dynamically relaxed clusters known within
$z<0.7$. (Our targets represent the most relaxed $\sim 20$ per cent of
clusters in this mass and redshift range.) We have, therefore, also
determined the best-fitting power-law parameters appropriate for the
most relaxed 20 per cent of simulated halos with $M_{\rm
vir}>2\times10^{14}h^{-1}$\Msun) in the \citet{Shaw05} study. (In
detail, this is done by selecting only those clusters with a
substructure fraction, as defined by those authors, $f_{\rm
s}<0.045$).  The resulting power-law fit parameters are $c_0=6.8$ and
$a=-0.16$. Although giving slightly better agreement with the
observations, this model still provides a poor description of the
Chandra data with $\chi^2=102.9$ for 30 DOF.

Physical motivation for the power-law model and its $(1+z)^{-1}$
redshift dependence is presented by \citet{Bullock01}. In their model
they assume that the ratio of the virial mass, $M_{\rm vir}$, to the
volume inside the scale radius, $V_{\rm s}$, is proportional to the
matter density $\rho_{\rm m}(z_{\rm coll})$ at the redshift, $z_{\rm
coll}$, when the cluster formed:

\begin{equation}
\rho_{\rm m} (z_{\rm coll})\propto \frac{M_{\rm vir}}{V_{\rm
s}}\propto\frac{\rho_{\rm m}\;r_{\rm vir}^3}{r_{\rm s}^3}=
\rho_{\rm m}\;c_{\rm vir}^3.
\end{equation}
Since the matter density $\rho_{\rm m}$ was larger in the past,
$\rho_{\rm m}\propto(1+z)^3$, the typical concentration parameter for
a halo of mass $M_{\rm vir}$ is smaller at higher redshifts, $c_{\rm
vir}\propto(1+z)^{-1}$.  \citet{Bullock01} associate each redshift
with a typical collapsing mass (defined as a fraction $F$ of the final
halo mass) through a critical value of the variance $\sigma(z,M)$ of
the density fluctuations. Since $\sigma(M)$ for a $\Lambda$CDM model
is a power-law for masses $M_{\rm vir}\lesssim 10^{13}\,M_{\odot}/F$,
this model yields a power-law for the mass-concentration relation (up
to this mass scale) with two free parameters. \citet{Bullock01}
suggest $F=0.01$ (although $F=0.001$ is also used in the more recent
literature). At the high-mass end, $M_{\rm vir}>10^{13}M_{\odot}/F$,
the model predicts a change in the power-law slope but, unfortunately,
also becomes unrealistic as the linear evolution of density
fluctuations stalls in the $\Lambda$CDM model, preventing high-mass
clusters from forming \citep{Bullock01,Kuhlen05}.

Although the \citet{Bullock01} model and the power law model of
\citet{Dolag04} are attractive in terms of their simplicity, the
limitations of such models for describing the detailed properties of
cluster halos should not be overlooked. Firstly, real clusters contain
X-ray emitting gas and stars, as well as dark matter. CDM-only simulations 
do not include cooling and feedback processes
which affect the baryonic mass components and modify the
overall mass distributions. Secondly, to date, studies of the
mass-concentration relation have included very few halos at the
largest mass range spanned by the Chandra data; e.g., \citet{Shaw05}
have only a single cluster with $M_{\rm vir}>10^{15}h^{-1}$\Msun.
Finally, \citet{Zhao03} argue that at the highest masses, the
\citet{Bullock01} model may over-predict the evolution of $c_{\rm vir}$
with redshift.

Motivated by such considerations, we have introduced additional
freedom into the power law model of \citet{Dolag04}. As well as having
$c_0$ and $a$ as free fit parameters, we also allow the redshift
evolution to evolve as $(1+z)^{-b}$, with $b$ free.

\begin{equation}
c=\frac{c_0}{(1+z)^b}
\left(\frac{M_{\rm vir}}{8\times 10^{14}\,h^{-1}M_{\odot}}\right)^{a}.
\label{plawfree}
\end{equation}
The results from a fit with this model, with $c_0$, $a$ and $b$ all
free, are shown by the solid line in Fig.~\ref{mvirc}. The model
provides a significantly improved description of the data with
$\chi^2=41.5$ for 29 degrees of freedom and best-fitting parameters
$c_0=7.55\pm0.90$, $a=-0.45\pm0.12$ and  $b=0.71\pm0.52$ (95 per cent
confidence limits; similar results are obtained from a Monte
Carlo analysis of the ($c_{\rm vir}|M_{\rm vir}$) data wherein 
the results for individual clusters are scattered according to their 
measurement errors). Although the normalization $c_0$
at $8\times 10^{14}\,h^{-1}M_{\odot}$
and the  redshift
evolution, $b$, are 
consistent with the \citet{Shaw05} simulations and \citet{Dolag04} model, 
the mass-concentration relation is noticeably steeper.

Fig.~\ref{cz} shows the variation of concentration parameter with
redshift.  At low redshifts, the fit to the \citet{Shaw05} simulated
data with the \citet{Dolag04} model ($c_0=6.47$, $a=-0.12$, $b=1$)
provides a reasonable match to the data (in agreement with the
conclusions drawn by \citealt{Pointecouteau05,Zhang06,
Vikhlinin06,Voigt06}).  At higher redshifts, however, the observed
$c_{\rm vir}$ values exceed the \citet{Shaw05} model predictions.  The
most striking feature of Fig.~\ref{cz} is an apparent absence of
redshift evolution in the concentration parameter. Indeed, when fixing
the slope parameter $a$ to the \citet{Shaw05} value of $a=-0.12$, a
fit with $c_0$ and $b$ free gives a redshift evolution parameter
$b=0.31\pm0.49$ (95 per cent confidence limits), consistent with no
evolution or even positive evolution of $c_{\rm vir}$ with $M_{\rm
vir}$. However, note that the $\chi^2$ for this fit is significantly
worse ($\chi^2=74.7$ for 30 DOF; $\Delta \chi^2=33.2$) than for the
fit in which the slope, $a$ of the mass-concentration relation is also
included as a free parameter (see above).  We conclude that any
analysis of redshift evolution in the mass-concentration relation
should (at least) explore the full parameter space discussed above,
with $c_0$, $a$ and $b$ included as free parameters.
  
Finally, we have estimated the systematic scatter that may be present
in the observed mass-concentration relation. This was carried out by
modifying the $\chi^2$ estimator to include an additional systematic
uncertainty, and increasing the size of this systematic uncertainty
until the reduced $\chi^2$ value became equal to unity. Based on the
fit with $c_0$, $a$ and $b$ included as free parameters, we estimate
an intrinsic systematic scatter in the data of $\Delta {\rm log}(c) \sim
0.1$, in good agreement with the predictions from simulations 
\citep[e.g., ][]{Bullock01,Wechsler02}.

In summary, the key results on the mass-concentration relation are 1)
the presence of a tight, observed mass-concentration relation for
massive, dynamically relaxed clusters. The
estimated intrinsic, systematic scatter
in this relation is
consistent with the predictions
from CDM simulations.  2) We observe a slope $a=-0.45\pm0.12$ (at 95 per
cent confidence) for the mass-concentration relation that is
significantly steeper than predicted by CDM simulations for lower 
mass halos ($a\sim
-0.12$; \citealt{Shaw05}).   3) The redshift 
evolution of the observed mass-concentration relation, $b=0.71\pm0.52$ 
at 95 per cent confidence, is consistent with
the value of 
$b=1$ in the \citet{Bullock01} model.

\begin{table}
\caption{Results on the concentration parameters $c_{\rm vir}=r^{\rm
total}_{\rm vir}/r^{\rm dark}_{\rm s}$ and virial masses $M_{\rm vir}$
(in units of $10^{14}\,h^{-1}M_{\odot}$) for the standard NFW model
(inner slope $\alpha=1$ fixed). Uncertainties are 68 per cent
confidence limits. If no error bar is quoted, the upper limit is
unbounded and the lower limit is zero.}
\label{m-c}
\begin{center}
\begin{tabular}{ l c c }
                       & $c_{\rm vir}$          &  $M_{\rm vir}$          \\
\hline		                                	  
Abell 1795             & $6.16^{+1.14}_{-1.14}$ & $7.58^{+1.92}_{-1.70}$  \\
Abell 2029             & $8.86^{+0.44}_{-0.50}$ & $7.06^{+0.69}_{-0.54}$  \\
Abell 478              & $5.15^{+0.45}_{-0.49}$ & $11.6^{+1.40}_{-1.80}$  \\
PKS0745-191            & $7.75^{+2.15}_{-1.41}$ & $10.4^{+4.70}_{-2.60}$  \\
Abell 1413             & $6.08^{+1.06}_{-1.06}$ & $9.76^{+2.54}_{-1.89}$  \\
Abell 2204             & $12.8^{+3.90}_{-2.83}$ & $6.86^{+2.56}_{-1.48}$  \\
Abell 383              & $5.08^{+0.55}_{-1.03}$ & $6.87^{+1.89}_{-1.85}$  \\
Abell 963              & $5.79^{+1.05}_{-1.28}$ & $7.00^{+2.45}_{-1.60}$  \\
RXJ0439.0+0521         & $8.20^{+2.00}_{-1.58}$ & $3.17^{+1.25}_{-0.87}$  \\
RXJ1504.1-0248         & $4.76^{+1.41}_{-1.47}$ & $15.1^{+9.40}_{-4.30}$  \\
RXJ2129.6+0005         & $5.23^{+3.12}_{-2.71}$ & $5.96^{+8.54}_{-2.70}$  \\
Abell 1835             & $4.18^{+0.63}_{-0.41}$ & $17.0^{+3.10}_{-3.40}$  \\
Abell 611              & $6.48^{+2.21}_{-2.14}$ & $6.18^{+3.82}_{-1.81}$  \\
Zwicky 3146            & $2.91^{+3.03}        $ & $24.8_{-13.6}$          \\
Abell 2537             & $6.15^{+2.96}_{-1.94}$ & $7.03^{+6.27}_{-2.81}$  \\
MS2137.3-2353          & $8.82^{+0.75}_{-0.79}$ & $3.89^{+0.58}_{-0.45}$  \\
MACSJ0242.6-2132       & $8.05^{+1.80}_{-1.13}$ & $3.91^{+1.24}_{-0.84}$  \\
MACSJ1427.6-2521       & $8.72^{+2.81}_{-2.78}$ & $2.39^{+1.47}_{-0.80}$  \\
MACSJ2229.8-2756       & $9.67^{+4.87}_{-3.39}$ & $2.48^{+1.62}_{-0.81}$  \\
MACSJ0947.2+7623       & $6.59^{+2.48}_{-1.90}$ & $9.30^{+6.00}_{-3.25}$  \\
MACSJ1931.8-2635       & $3.59^{+2.64}_{-2.38}$ & $11.5_{-4.60}$          \\
MACSJ1115.8+0129       & $2.22^{+2.80}        $ & $34.2_{-24.6}$          \\
MACSJ1532.9+3021       & $5.84^{+1.82}_{-1.63}$ & $7.73^{+4.27}_{-2.25}$  \\
MACSJ0011.7-1523       & $3.73^{+1.12}_{-0.80}$ & $8.27^{+3.53}_{-2.59}$  \\
MACSJ1720.3+3536       & $5.14^{+1.66}_{-1.05}$ & $6.91^{+3.49}_{-2.06}$  \\
MACSJ0429.6-0253       & $9.36^{+2.01}_{-1.37}$ & $3.15^{+0.91}_{-0.75}$  \\
MACSJ0159.8-0849       & $6.12^{+1.15}_{-1.53}$ & $10.5^{+3.50}_{-3.27}$  \\
MACSJ0329.7-0212       & $5.74^{+0.95}_{-1.04}$ & $5.67^{+1.75}_{-1.25}$  \\
RXJ1347.5-1144         & $5.51^{+0.90}_{-0.30}$ & $23.5^{+6.80}_{-4.40}$  \\
3C295                  & $9.30^{+1.31}_{-1.13}$ & $2.95^{+0.62}_{-0.48}$  \\
MACSJ1621.6+3810       & $7.11^{+3.75}_{-2.33}$ & $5.92^{+4.18}_{-2.26}$  \\
MACSJ1311.0-0311       & $5.35^{+1.81}_{-1.31}$ & $5.41^{+3.02}_{-1.66}$  \\
MACSJ1423.8+2404       & $9.20^{+0.83}_{-1.04}$ & $4.52^{+0.79}_{-0.64}$  \\
MACSJ0744.9+3927       & $5.04^{+1.92}_{-1.20}$ & $7.35^{+4.35}_{-2.12}$  \\
\hline
\end{tabular}
\end{center}
\end{table}

\begin{figure}
\begin{center}
\resizebox{\columnwidth}{!}{\includegraphics{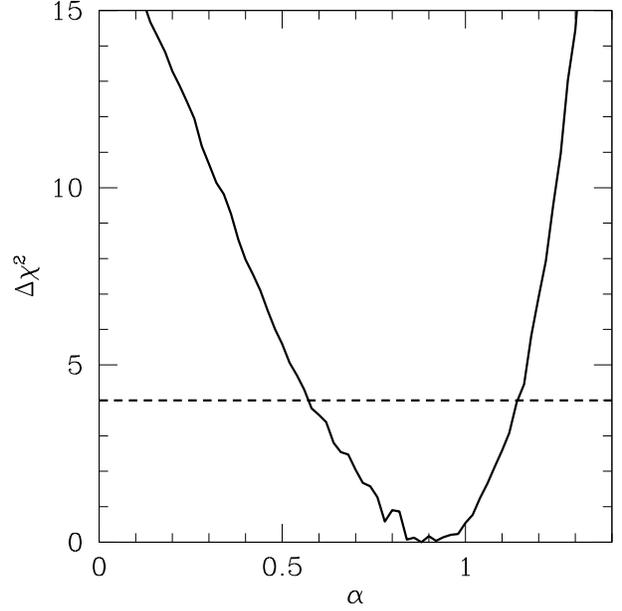}}
\end{center}
\caption{The summed $\chi^2$ values as a function of the inner density
slope $\alpha$ for the 27 clusters for which the generalized NFW model
provides a reasonable fit to the data. The dashed line indicates the
$2\sigma$ confidence limits. The overall best fit is obtained for
$\alpha=0.88^{+0.26}_{-0.31}$ (95 per cent confidence limits).}
\label{gnfwsum}
\end{figure}

\subsection{The inner slope of dark matter profiles: generalized NFW models}

For the final stage of our analysis, we have included the inner slope
of the dark matter profile ($\alpha$, in the generalized NFW model) as
an additional free parameter in the fits with method 2.
(Separate components to model the mass contributions from the X-ray emitting
gas and the optically luminous matter in the central, 
dominant galaxies were included in the fits.)

The results on the inner mass profiles are summarized
in Table~\ref{table:dm}.
In terms of the goodness of fit, 27 out of 34 
clusters give $\chi^2$/DOF values with a probability 
of $0.05$ or better.
Of these 27 clusters, 21 have an
inner slope consistent with unity at 68 per cent
confidence. 

In order to obtain a combined result on the inner dark matter
profile, we have summed together the results on $\chi^2$ 
as a function of $\alpha$ for the 27 clusters for 
which the generalized NFW model provides a reasonable 
description of the data. The results are shown in 
Fig.~\ref{gnfwsum}. (Note that we have subtracted the 
overall minimum summed $\chi^2$, of 75.37 for 72 DOF; 127
temperature measurements, 2 free parameters for 27 
clusters and one free slope parameter). 

From the summed $\chi^2$ data we obtain a best-fitting 
inner slope of $\alpha=0.88^{+0.26}_{-0.31}$, where the 
uncertainties are 95 per cent confidence limits.
We conclude that our combined result on the inner
dark matter density slope is consistent 
with the range of values predicted by CDM simulations.

\section{Discussion}

\subsection{Previous results on the inner dark matter density slopes}
\label{section:slopes}

Several of the clusters in the present sample have been the subjects
of previous work that has examined the issue of the inner dark matter
density slope. MS2137.3-2353 and Abell 963 were studied by
\citet{Sand02,Sand04} using a combination of strong gravitational
lensing data and optical velocity dispersion measurements for the
dominant cluster galaxies. These authors measured surprisingly flat
inner slopes for MS2137.3-2353, Abell 963 (which are formally
consistent with the findings from the Chandra data presented in
Table~\ref{table:dm}) and four other clusters, and concluded that
their results were inconsistent with the standard NFW model
($\alpha=1.0$) at 99 per cent confidence.

Later work by \citet{Bartelmann04} and \citet{Dalal03} cautioned that
allowing for deviations from axial symmetry in the strong lensing
analysis, using elliptical rather than spherical mass models (as
motivated by the data) will modify the constraints on the inner dark
matter profiles. These authors concluded that the strong lensing data
of \citet{Sand02,Sand04} remain consistent with CDM models, once such
effects are taken into account. The X-ray results presented here are
also consistent with the standard NFW model ($\alpha=1.0$) at 95 per
cent confidence.  It is important to note that X-ray data do not
suffer in the same way from uncertainties regarding axial
symmetry. For the regular, apparently relaxed clusters in the present
sample, the mass models determined from the X-ray data under the
assumption of spherical symmetry are unlikely to be affected by
asphericity effects by more than a few per cent
\citep{Piffaretti03,Gavazzi05}.

Using Chandra data for Abell 2029 (the first $\sim 20$ks observation
only) and parameterized models for the gas density and temperature
profiles, \citet{Lewis03} obtained $\alpha=1.19\pm0.04$ for the inner
(total) mass profile. We measure $\alpha=1.16^{+0.22}_{-0.34}$ for the dark
matter at 68 per cent confidence using 94ks of clean data. 
(For the total mass, a fit with a generalized NFW model gives 
$\alpha=1.16^{+0.26}_{-0.22}$). Thus, the present study and 
\citet{Lewis03} obtain similar best-fit results, although the
statistical uncertainties reported here are significantly larger, 
despite being based on more data. In part, this
highlights the effects that priors, in the form of parameterized
models for the gas density and temperature profiles, can have
on the analysis.  \citet[see also \citealt{Buote04}]{Zappacosta06}
present a detailed study of the nearby, intermediate temperature
cluster Abell 2589, for which they also model the dark matter, X-ray
emitting gas and stellar mass associated with the cD galaxy
separately. They conclude that the standard NFW model with $\alpha=1$
provides a good description of the dark matter and total mass
distributions in the cluster.
 
\citet{Arabadjis04} studied Chandra data for Abell 1835, Abell 2029,
Abell 2204, Zwicky 3146 and MS 2137.3-2353.  Fig. 4 of that paper
indicates best-fitting values of $\alpha\sim 1.85$ for Abell 2029,
$\alpha\sim 1.8$ for Abell 2204, $\alpha\sim 0.9$ for Abell
1835, $\alpha\sim 1.7$ for Zwicky 3146 and $\alpha=1.6\pm0.2$ for
MS 2137.3-2353 (with 68 per cent confidence limits of
$\sim 10-20$ per cent). In general, we find shallower inner dark
matter slopes than \citet{Arabadjis04}. Our statistical uncertainties
are also larger in some cases.

\citet{Voigt06} present results on the inner slopes for 12
clusters, 8 of which are in common with the present study. These
authors use a similar non-parametric spectral deprojection technique
to measure the deprojected X-ray temperatures in the clusters. Their
results on the central slope are in general agreement with those
presented here.

\subsection{On the robustness of the central galaxy model}
\label{stars}

\citet{Sand02} present a detailed analysis of the optical properties
of the dominant galaxy in MS2137.3-2353 (see also
\citealt{Gavazzi05}). These authors fit a de Vaucouleurs profile to
the galaxy surface brightness, yielding an effective radius $R_{\rm
e}=5.02\pm0.50$ arcsec and a total V-band luminosity $L_{\rm
V}=4.2\times10^{11}\,L_{\odot}$.  Using the
redshift-dependent V-band mass-to-light ratio of \citep{Fukugita98}
and assuming a galaxy formation redshift $\sim 2$, we obtain a V-band
mass-to-light ratio for the dominant galaxy in MS2137.3-2353 of
$M/L_V=2.75$ and a total stellar mass of
$1.14\times10^{12}$\Msun. (The stellar mass dominates the total mass
within $10$ kpc of the center of the cluster. Note that
\citet{Sand02} measure an optical velocity dispersion for the dominant
galaxy in MS2137.3-2353 of $\sigma \sim 275$ km/s, from which they
infer $M/L_V\sim 3.1$; see also \citealt{Gavazzi05} for a slightly
lower $M/L_V$ value). We have adopted these parameters as a template
to approximate the mass contribution from stars in the central
galaxies of all clusters in the sample, employing either a Jaffe model
or de Vaucouleurs model.

In principle, we might expect the results on the inner dark matter
density slopes to be sensitive to the choice of parameters
used to describe the central stellar mass components.  Here, the main
systematic uncertainty lies with the assumption that the central
stellar mass distribution in MS2137.3-2353 provides a reasonable model for
other galaxy clusters in the sample. The assumption of a constant
central stellar mass is well motivated; K-band observations of
dominant cluster galaxies in X-ray luminous clusters \citep{Brough02}
show little variation in total stellar mass over the redshift range
$0<z<1$, with scatter at the level of $20-30$ per cent.

\begin{figure}
\begin{center}
\resizebox{\columnwidth}{!}{\includegraphics{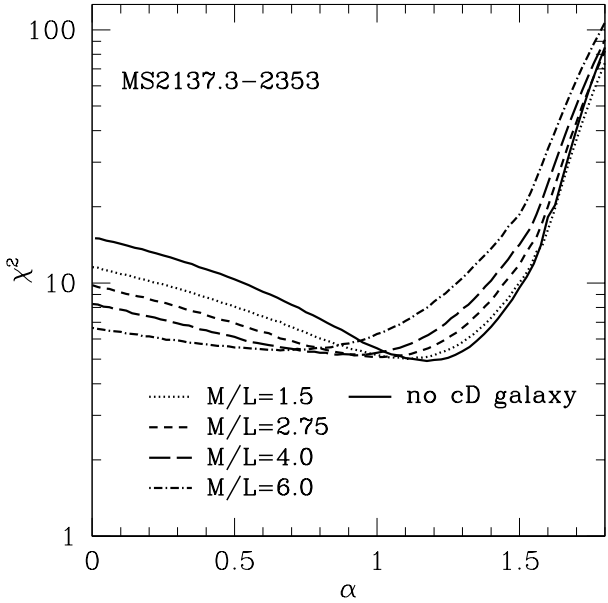}}
\end{center}
\caption{Goodness of fit for the inner slopes $\alpha$ for the galaxy
  cluster MS2137.3-2353. The mass model includes a Jaffe component for
  the luminous matter of the cD galaxy and a generalized NFW component
  for the dark matter halo.}
\label{starsfig}
\end{figure}

To estimate of the effect of departures from a constant central
stellar mass on the measured inner dark matter slopes, we have
re-analysed the Chandra data for MS2137.3-2353 varying the $M/L_V$
ratio over the range $M/L_V=0-6$ (and thereby changing the central
stellar mass by $\sim \pm 100$ per cent).  In each case, we have
re-determined the constraints on the inner slope $\alpha$. The results
are shown in Fig.~\ref{starsfig}.  Note that MS2137.3-2353 is one of
the least massive clusters in the sample, and so changes in $\alpha$
with central stellar mass are likely to be larger than for most other
clusters in the sample.  For $M/L_V=1.5$, we measure
$\alpha=1.10^{+0.22}_{-0.30}$ at 68 per cent confidence. For $M/L_V=4$,
we measure $\alpha=0.94^{+0.24}_{-0.46}$. The tendency is for higher
$M/L_V$ ratios to give slightly shallower inner dark matter slopes.
These results can be compared to the best-fitting result for MS2137.3-2353
with $M/L_V=2.75$ of $\alpha=1.00^{+0.25}_{-0.35}$.  We conclude that the
results on the inner dark matter slope for MS2137.3-2353 are robust
against changes in the central stellar mass by $\pm 50$ per cent.

Finally, we note that extending the analysis to model the mass
components associated with stars external to central galaxies
separately should have a negligible effect on the results. In total,
stars contribute only $\sim 2$ per cent of the total cluster mass
\citep[e.g., ][]{Lin04,Fukugita98}
 
In conclusion, the results presented here are robust against
systematic uncertainties associated with measuring the stellar mass
contribution in the clusters.

\subsection{Residual systematic uncertainties}
\label{systematics}

The clusters studied here are the most massive, dynamically relaxed
clusters known and are, therefore, the systems for which the
assumption of hydrostatic equilibrium should be most robust.  Although,
as discussed above, geometric effects and uncertainties associated
with separating the dark matter and baryonic matter components are
unlikely to impact on the results significantly, some systematic
uncertainties remain.

In the first case, it remains possible that small levels of
non-thermal pressure support due to \eg gas motions, cosmic rays or 
magnetic fields could be present in the X-ray emitting gas and bias the 
measured masses low. However, the relaxed nature of the clusters 
argues that large, unaccounted for, bulk and/or turbulent 
motions are unlikely to be present and that the mass 
measurements should be recovered 
to better than 10-20 per cent accuracy \citep{Nagai06,Rasia06}.
Galaxy clusters are known to contain magnetic fields \citep[e.g., ][]{Carilli02}
The issue of magnetic pressure support and its effect on X-ray
mass measurements has been studied by \citet{Dolag00} who
showed that for relaxed clusters, magnetic pressure support is
unlikely to bias mass measurements significantly, even in the central
regions. For individual clusters, effects no larger than 10-20 per
cent are expected. The similarity of the mass results obtained using 
method 1, where we measure the total mass and method 2, where we
model the dark and baryonic
components separately, also argues that any non-thermal
pressure component, at the $\sim 10-15$ per cent level, distributed in
a similar manner to the dark or baryonic matter, is unlikely to have a
significant effect on the conclusions.  A program to determine the
maximum size of non-thermal pressure in the clusters, using a
combination of Chandra X-ray data and wide field weak gravitational
lensing observations, is underway (Donovan \etal, in preparation).

It should also be recognized that although we quote results on the
mass-concentration relation appropriate for the virial radii in the
clusters, thereby allowing an easy comparison with the predictions
from simulations, the X-ray data only extend to about $r_{2500}$ or
approximately a quarter to a third of the virial radius in most
clusters. Some systematic uncertainty is associated with extrapolating
the allowed range of NFW mass models to these larger radii.

Finally, we reiterate that the simulations used to predict the
mass-concentration relation and central dark matter slopes
\citep[e.g., ][]{Navarro95,Navarro97,Diemand04,Shaw05} are CDM
only. In detail, these predictions may be modified by future numerical 
work that includes the X-ray emitting gas and stars, and realistic
baryonic physics (cooling, star formation, AGN feedback). For the
most accurate comparison with the data presented here, such
simulations should contain a sufficiently large number of massive,
relaxed clusters and be normalized to match the observed X-ray (e.g.,
temperature profiles, virial relations, X-ray gas mass fraction;
\citealt{Allen01b,Allen04,Vikhlinin06}) and optical
\citep{Croton06,Bower06} properties.

\section{Summary}

We have used the Chandra X-ray Observatory to measure the dark matter
and total mass profiles for a sample of 34 massive, dynamically
relaxed galaxy clusters. Our analysis has employed a non-parametric,
spherical deprojection technique that minimizes the need for 
priors associated with parameterized models for the X-ray gas density
and/or temperature profiles. This allows a direct 
assessment of the goodness of fit to the Chandra data provided by a
variety of simple mass models as well as an accurate determination of
statistical uncertainties on fit parameters.

We have shown that the NFW model, which is motivated by CDM
simulations, provides a good description of the total mass and dark
matter distributions in the majority of clusters. In contrast, the
singular isothermal sphere model can, in almost every case, 
be firmly ruled out. Combining
the results for all clusters for which the NFW model provides an
acceptable description of the dark matter profiles, we obtain a
best-fitting result on the inner slope of the dark matter density profile
in the clusters, $\alpha=0.88^{+0.15}_{-0.11}$ (68 per cent confidence
limits).

We observe a well-defined mass-concentration relation for the clusters
with an
intrinsic scatter in good agreement with the
predictions from simulations. The slope of the mass-concentration
relation, $c\propto M_{\rm vir}^a/(1+z)^b$ with $a=-0.45\pm0.12$ at 95
per cent confidence, is significantly steeper than the value of $a\sim
-0.1$ predicted by CDM simulations for lower mass halos. The redshift
evolution, $b=0.71\pm0.52$ at 95 per cent confidence, is  
consistent with the value $b\sim 1$ predicted by
those simulations.

After this work was preprinted on astro-ph, \citet{Buote06}
preprinted a paper in which they discuss the mass-concentration
relation for 39 systems with masses in the range $6\times 10^{12}
M_{\odot}$ to $2\times10^{15} M_{\odot}$. The results for higher mass
systems are drawn from the previous studies of \citet{Pointecouteau05}
and \citet{Vikhlinin06} and are in good overall agreement with the
present work.

\section*{Acknowledgments}

We thank Laurie Shaw for kindly providing the simulated data from
\citet{Shaw05} and Adam Mantz and Glenn Morris for helpful
discussions. We thank our collaborators in the ongoing cluster cosmology
work, especially H. Ebeling for his heroic efforts in
compiling the MACS sample. We are grateful to the developers of the
GNU Octave numerical computation language and the GNU Scientific
Library for their ongoing efforts.  This work was supported in part by
the U.S. Department of Energy under contract number DE-AC02-76SF00515
and by the National Aeronautics and Space Administration through
Chandra Award Number DD5-6031X issued by the Chandra X-ray Observatory
Center, which is operated by the Smithsonian Astrophysical Observatory
for and on behalf of the National Aeronautics and Space Administration
under contract NAS8-03060.



\bsp

\label{lastpage}

\end{document}